\documentclass[%
prd%
,twocolumn%
,amssymb]{revtex4}
\usepackage{latexsym}
\usepackage{graphicx}
\newcommand{\ovl}[1]{\overline{#1}}

\begin{document}

\title{
Deconfinement transition and string tensions
in SU(4) Yang-Mills Theory
}

\author{Matthew Wingate}
\affiliation{RIKEN BNL Research Center,
Brookhaven National Laboratory, Upton, NY 11973, USA}
\author{Shigemi Ohta}
\affiliation{Institute for Particle and Nuclear Studies, KEK,
Tsukuba, Ibaraki 305-0801, Japan}
\affiliation{and RIKEN BNL Research Center,
Brookhaven National Laboratory, Upton, NY 11973, USA}

\date{\today}

\begin{abstract}\noindent
We present results from numerical lattice calculations of SU(4) 
Yang-Mills theory.  This work has two goals: to determine the
order of the finite temperature deconfinement transition
on an $N_t = 6$ lattice and to study the string tensions
between static charges in the irreducible representations of
SU(4).  Motivated by Pisarski and Tytgat's argument that a
second-order SU($\infty$) deconfinement transition would 
explain some features of the SU(3) and QCD transitions,
we confirm older results on a coarser, $N_t = 4$,
lattice.  We see a clear two-phase coexistence signal in
the order parameter, characteristic of
a first-order transition, at
$8/g^2 = 10.79$ on a $6\times 20^3$ lattice, on which we
also compute a latent heat of $\Delta\epsilon\approx 0.6 
\epsilon_{\rm SB}$.
Computing Polyakov loop correlation functions 
we calculate the string tension 
at finite temperature in the confined phase
between fundamental charges, $\sigma_1$, between diquark 
charges, $\sigma_2$,
and between adjoint charges $\sigma_4$.  
%We find that $1 < \sigma_2/\sigma_1 < 2$, and we see
%{\it evidence} for adjoint string breaking, $\sigma_4 = 0$.
We find that $1 < \sigma_2/\sigma_1 < 2$, and our result
for the adjoint string tension $\sigma_4$ is consistent
with string breaking.
\end{abstract}

\maketitle
\newpage
%%%%%

\section{Introduction}
\label{sec:intro}
It is well-established that the dynamics of the strong 
force are described by nonabelian gauge theory
%as first constructed by Yang and Mills~\cite{ref:YANG_MILLS},
with an internal SU(3) symmetry and matter in the
fundamental representation.  Although the full 
theory of QCD contains dynamical quarks, study of SU(3) pure
Yang-Mills theory~\cite{ref:YANG_MILLS} provides useful physical information.  
For example, in the valence or quenched approximation where
gauge field configurations are generated without including the
fermion determinant in the partition function, the computed
light hadron spectrum differs from the 
experimentally measured spectrum at the 10\% level~\cite{ref:CPPACS_LAT98}.
This is convenient
since Monte Carlo calculations require enormous computational effort
to include dynamical quark effects, so many studies
are done in the quenched approximation in the interest of practicality.  
In the present paper we are interested in the phase diagram of QCD
in the temperature--quark mass plane, and so the study of
pure Yang-Mills theory covers the $m_q = \infty$ line.

The confinement--deconfinement transition of QCD 
at high temperature ($T \approx 100{\rm -} 300$ MeV) 
has been studied with and
without dynamical quarks and depends strongly on the number
of light quark flavors and their masses.  Pure gauge theory
is recovered in the limit of infinite quark masses.
In this limit the order parameter of the deconfinement transition
is the Polyakov loop \cite{ref:POLY_SUSS},
\begin{equation}
L(\vec{x}) = {1\over N_c} ~{\rm Tr}~
\prod_{t=1}^{N_t} U_0(\vec{x},t).
\label{eq:ploop}
\end{equation}
In the confined phase $\langle L\rangle_{\vec x} = 0$, but in the
deconfined phase the Polyakov loop acquires a nonzero expectation
value, spontaneously breaking the global Z($N_c$) center 
symmetry.  Since a Z(3) symmetry admits a cubic term in the
effective potential, it drives a first-order transition; such is
not the case for $N_c > 3$~\cite{ref:SVET_YAFFE}.

\begin{figure}
\includegraphics[width=2.5in,bb=75 126 498 540]{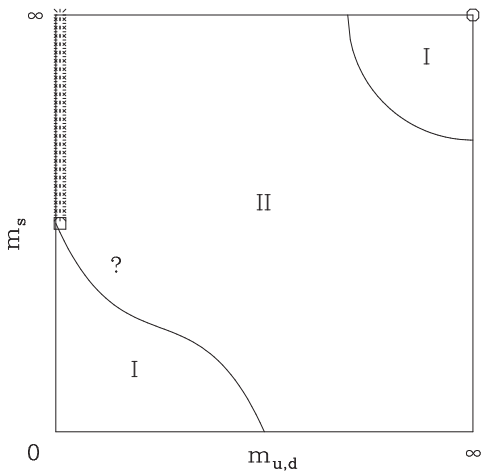}
\caption{Columbia diagram \cite{ref:CU_2P1}, showing the nature of the 
2+1-flavor finite temperature transition for different values of 
$m_{u,d}$ and $m_s$.  Regions labeled I have a first-order transition,
while the region II has only a crossover.  For $m_u=m_d=0$ there is
a tri-critical point (square) above which the chiral transition is
second-order, denoted by asterisks.  
The pure glue limit is indicated by the octagon
at $(\infty,\infty)$.  The question mark indicates the physical
$(m_{u,d},m_s)$ according to calculations with staggered fermions
\cite{ref:KS_2P1}.  (Ref.\ \cite{ref:WIL_2P1} suggests that the 
physical point lies lower, in the Region I, when Wilson fermions are used.)
}
\label{fig:columbia}
\end{figure}

For $N_f$ flavors of massless quarks, the QCD Lagrangian
has a global SU$(N_f)_L\otimes$SU$(N_f)_R$ chiral symmetry.
At zero temperature this symmetry is spontaneously broken
and the pions are the massless Goldstone bosons, but at some
finite temperature $T_\chi$ the chiral symmetry is restored.
Universality arguments suggest that the transition should
be first-order for $N_f\ge 3$~\cite{ref:PIS_WIL}, while
a second-order transition for $N_f=2$ is not ruled
out~\cite{ref:WIL_RAJ} (in fact a second-order transition
is supported by lattice studies~\cite{ref:KARSCH_LAT99}).
In nature the strange quark mass, $m_s$, is roughly 25 times larger
than the average of up and down quark masses, $m_{u,d}=(m_u+m_d)/2$. 
According to lattice calculations, the order of this ``2+1'' flavor phase
transition depends on the strange mass.  As $m_s$ is increased from
$m_{u,d}$, the first-order phase transition weakens into a 
crossover~\cite{ref:CU_2P1}.  In Fig.~\ref{fig:columbia} we
reproduce the ``Columbia'' phase diagram which
shows the order of the transition for different regions in $(m_{u,d},m_s)$
plane.

Recently Pisarski and Tytgat~\cite{ref:PISARSKI_TYTGAT} 
argued that the Columbia diagram is hard to understand in
light of intuitive large-$N_c$ arguments.  They point out
that since anomaly effects are suppressed by $1/N_c$, the
contribution of chiral symmetry restoration to the free 
energy is $O(N_c)$ while the change in the free energy due
to deconfinement is $O(N_c^2)$.  So, in the large-$N_c$ limit
a first-order deconfinement transition should be robust for
any quark mass.  Thus, if the the first-order transition 
of SU(3) is a general feature of SU($N_c$) it is hard to understand
why it disappears as the quark masses
increase away from zero.
One resolution of this conflict, they propose,
is that $N_c = 3$ is special due to the cubic term in the
effective potential, and that the general SU($N_c$) deconfinement
transition is second order.  

\begin{figure}
\includegraphics[width=2.5in,bb=88 138 495 538]{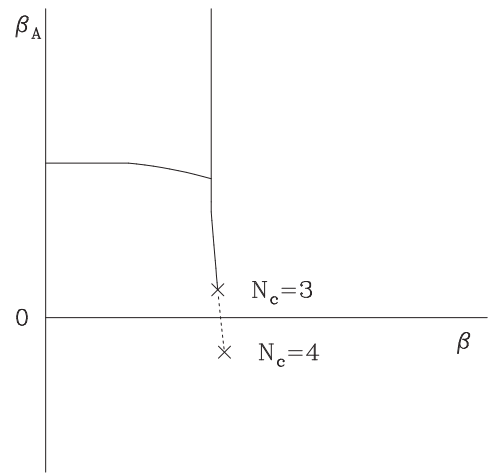}
\caption{Phase diagram of fundamental-adjoint lattice action
showing the lattice-induced bulk transition.
The transition line crosses the fundamental axis for $N_c > 3$.
}
\label{fig:fund_adj}
\end{figure}

Yang-Mills theory with $N_c$ colors, SU($N_c$) pure-gauge theory,
has been a topic of exploration
for lattice Monte Carlo study since the first days of the
field~\cite{ref:CREUTZ_SU2}.  A first-order phase transition
in the average energy was observed on symmetric lattices with volumes
between $3^4$ and $6^4$ for $N_c = 2-5$
\cite{ref:BALIAN,ref:CREUTZ_SU2,ref:CREUTZ_SU3,ref:BCM,ref:CREUTZ_SU5}.
%CHECK FOR EARLIER REF.\ BY BALIAN, DROUFFE, ITZYKSON.
Refs.\ \cite{ref:BHANOT_CREUTZ,ref:BKL,ref:DHN}
demonstrated this transition
is the consequence of a lattice-induced critical line separating 
the strong and weak coupling regimes.  Specifically, they added
to the usual fundamental single-plaquette action
\begin{equation}
S_f ~=~ -\frac{\beta}{6} \sum_x \sum_{\mu,\nu:\mu<\nu}
 {\rm Re~ Tr}_f ~P_{\mu\nu}(x)
\label{eqn:action}
\end{equation}
the adjoint action
\begin{equation}
S_A ~=~ -\frac{\beta_A}{6} \sum_x \sum_{\mu,\nu:\mu<\nu}
 {\rm Re~ Tr}_A ~P_{\mu\nu}(x),
\end{equation}
where $P_{\mu\nu}(x)$ is the plaquette 
$U^\dagger_\nu(x)U^\dagger_\mu(x+\hat\nu)U_\nu(x+\hat\mu)U_\mu(x)$ and
Tr$_f$ and Tr$_A$ are traces in the fundamental and adjoint
representations, respectively; $\mu$ and $\nu$ are space--time
(or space--temperature) indices.
Figure~\ref{fig:fund_adj}
shows the resulting phase diagram for this mixed action for
SU(3) and SU(4).  A first-order transition line separates
the strong and weak coupling regimes of the fundamental coupling
$\beta$.  For $N_c \le 3$ this line ends in a critical point
before crossing the $\beta_A=0$ axis, but does cross this axis
for $N_c > 3$.  Also displayed is the $\beta \approx 0$ transition
line corresponding to the transition in SO($N_c^2-1$) gauge theory.

This lattice-induced ($4d$) bulk transition, while interesting, can
obscure the physical ($3d$) finite temperature transition of
interest here.  If the two transitions are nearby in parameter
space, the first-order nature of the bulk transition would
have a non-trivial effect on the confinement-deconfinement
transition.  The bulk transition was found to occur near
$\beta=10.2$ (with $\beta_A=0$)~\cite{ref:BHANOT_CREUTZ}.
The past finite temperature studies of SU(4) 
\cite{ref:GOCKSCH_OKAWA,ref:GREEN_KARSCH, ref:BAT_SVET,ref:WHEAT_GROSS}
addressed this issue to varying degrees.  For example,
Ref.~\cite{ref:BAT_SVET} studied the deconfinement transition
along the line $\beta_A = -\beta/2$, hoping to avoid crossing
the bulk transition line.  Ref.~\cite{ref:WHEAT_GROSS} 
found evidence on small volumes and $\approx 1000$ 
Monte Carlo evolution sweeps for deconfinement transitions for both
$N_t =4$ and $N_t=5$ around $\beta=10.5$ and $\beta=10.6$
(with $\beta_A=0$).  All these studies concluded that
the SU(4) deconfinement transition was first-order.
However, given the alluring explanation for the Columbia
diagram, we felt the time was ripe to revisit finite temperature
SU(4) numerically.

Another topic which we address in this work is
the tension of confining strings which carry $k\in\{1, ...,
N_c\}$  units of flux.  Only for $N_c>3$ can one find different strings
with unequal tensions.  With these finite-temperature calculations we find
the ratio of diquark ($\sigma_{k=2}$) to fundamental ($\sigma_{k=1}$)
string tensions to be in the range $1< \sigma_{k=2}/\sigma_{k=1} < 2$.  As
pointed out in Ref.~\cite{ref:STRASSLER},  these types of computations
may test dualities between gauge theories and string theories.  String
tensions can be computed on the lattice, in broken
supersymmetric gauge theory,
and in M theory versions of QCD and supersymmetric QCD.
Our result for $\sigma_2/\sigma_1$ indicates that in
SU(4) Yang-Mills flux tubes attract
each other as expected from SUSY Yang-Mills and M theory \cite{ref:STRASSLER}
and proved in standard Yang-Mills \cite{ref:CREUTZ_PRIVATE}.

In the next Section we describe some details of our calculations.
Section~\ref{sec:trans} gives the results for the deconfinement
transition, and Section~\ref{sec:string} shows our calculation
of the string tensions.  Finally we summarize our
results in Section~\ref{sec:concl}.

%%%%%%
\section{Computation}
\label{sec:simul}

Our calculations of SU(4) Yang-Mills theory do not differ
significantly from standard SU(3) calculations.  We use the fundamental
single-plaquette action, Equation~(\ref{eqn:action}), with \(\beta =
2N_c/g^2 = 8/g^2\).  Our production code is a minimally modified
version of the MILC code~\cite{ref:MILC_CODE}.  Our algorithm for
evolving the gauge fields is a mixed overrelaxation/heatbath procedure:
in one Monte Carlo ``sweep'' we perform 10 microcanonical overrelaxation
steps followed by one  Kennedy-Pendleton~\cite{ref:KEN_PEND} heatbath
step.  Each sweep we compute the average plaquette  and fundamental
Polyakov loop.  We generate at least 1000 Monte Carlo sweeps at each
\(\beta\), with 10000 to 20000 sweeps around \(\beta_c\).
An independent Metropolis code was written from scratch
for SU(4) to check this MILC-derived SU(4) code.

For those values of the coupling $\beta$ where we want to calculate
the string tension, we compute correlation functions of Polyakov
loops in the irreducible representations of SU(4)
\begin{equation}
C_i(r)=\langle L_i(\vec{x}) L_i^*(\vec{x}+\vec{r}) \rangle_{\vec{x}},
\label{eq:corrfn}
\end{equation}
where $i$ ={\bf 4}, {\bf 6}, {\bf 10} and {\bf 15} and
the trace in Eq.~(\ref{eq:ploop}) is $i$-dimensional.  As is well known,
the diquark {\bf 6} and {\bf 10} representations are obtained by
antisymmetrizing and symmetrizing two fundamental {\bf 4}
representations, and the adjoint {\bf 15} by inserting SU(4) Gell-Mann
matrices.  We use the Parisi-Petronzio-Rapuano multihit variance reduction
method~\cite{ref:PPR} to reduce noise.
Polyakov loop correlation functions are computed every tenth Monte
Carlo sweep.  We investigate autocorrelations by including only every
\(n\)-th configuration, where \(n = 1\), 5, and 10.
We have a total of 2800 measurements for the 
calculations on a \(6\times16^3\) lattice and 1900 measurements
for the calculations on a \(8\times12^3\) lattice.

%%%%%%
\section{Deconfinement transition}
\label{sec:trans}

\begin{figure}
\includegraphics[width=3.2in,bb=14 -3 677 420]{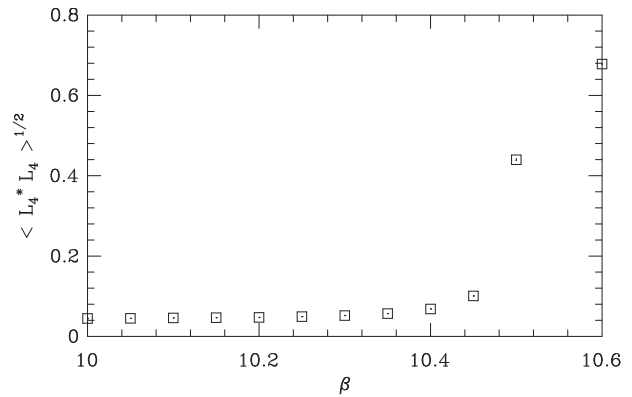}
\caption{Magnitude of the Polyakov loop vs.\ $\beta$ on a 
$4\times 8^3$ lattice.}
\label{fig:plmag_84}
\end{figure}

\begin{figure}
\includegraphics[width=3.2in,bb=21 -3 677 420]{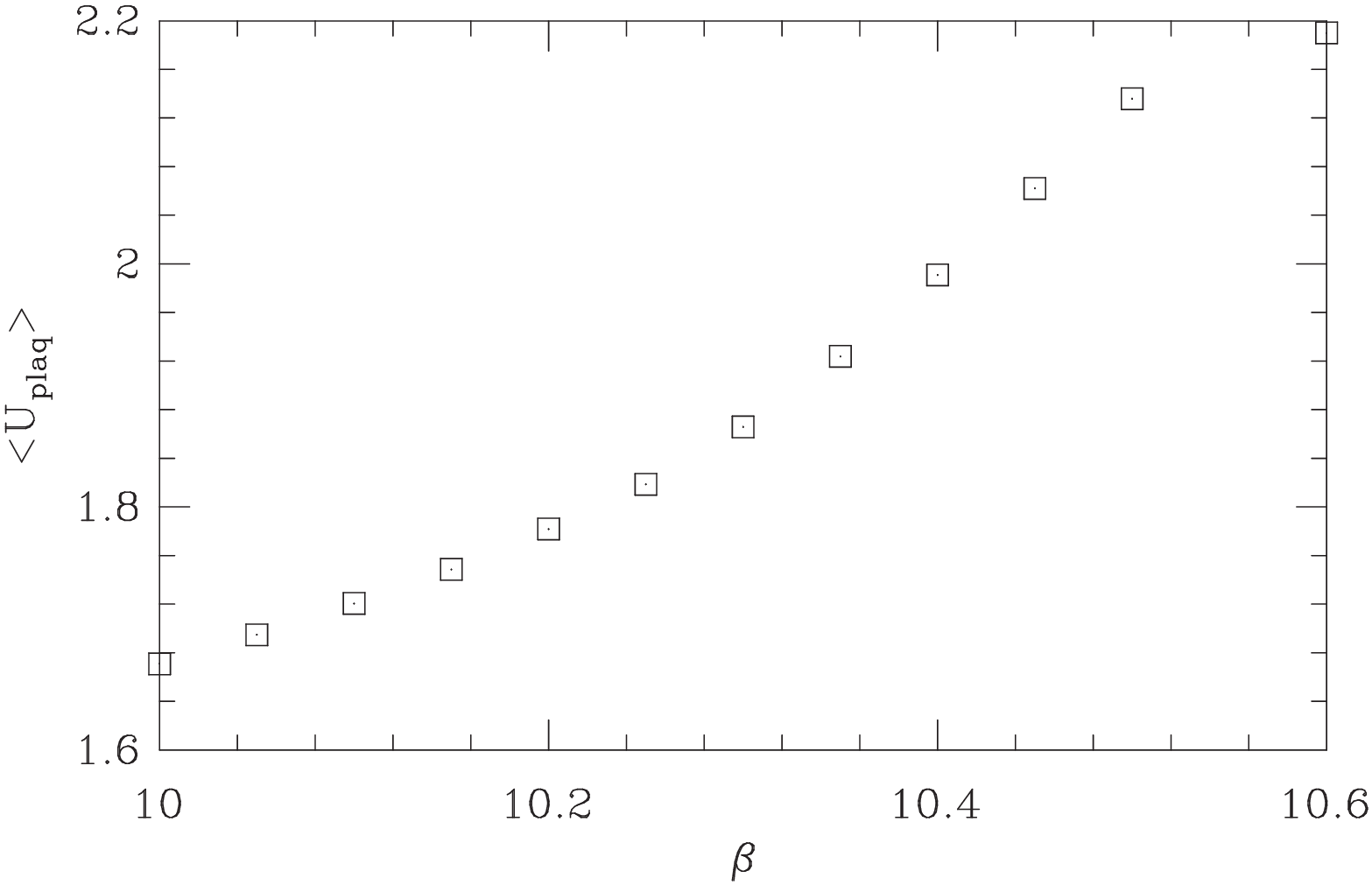}
\caption{Plaquette vs.\ $\beta$ on $4\times 8^3$ lattice.
The normalization is such that $\langle U_\mathrm{plaq}\rangle = N_c$ 
in the free theory. }
\label{fig:plaq_84}
\end{figure}

In order to make contact with previous finite temperature
SU(4) calculations, we compute thermodynamic observables on
a $4\times 8^3$ lattice for values of $\beta = 8/g^2$ between
10.0 and 10.6.  
We find a rapid change in $\langle|L|\rangle$ between
$\beta = 10.45$ and $\beta=10.5$ (see Fig.~\ref{fig:plmag_84}), 
in agreement with
Refs.~\cite{ref:GOCKSCH_OKAWA,ref:WHEAT_GROSS}.
Since the
plaquette is also increasing in that region (Fig.~\ref{fig:plaq_84}), 
one might worry that the bulk transition is affecting
the deconfinement transition.  Therefore, we did not pursue
confirming the order of the deconfinement transition
with $N_t=4$, and instead focus on $N_t=6$ where the
bulk and deconfinement transitions should be further 
separated.

%$12^3\times6$ paragraph:  Large change in plaquette at same value
%of $\beta$.  Deconfinement transition moved up in $\beta$, away
%from bulk transition.  Still no strong evidence for first-order
%transition.

\begin{figure}
\includegraphics[width=3.2in,bb=14 -3 658 420]{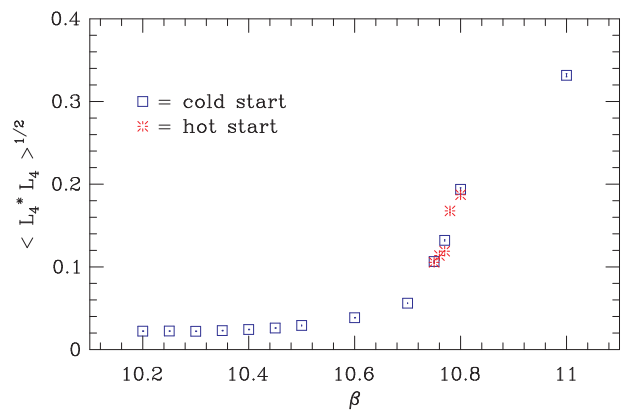}
\caption{Magnitude of the fundamental Polyakov loop vs.\ $\beta$ on a
$6\times 12^3$ lattice.
}
\label{fig:plmag_126}
\end{figure}
\begin{figure}
\includegraphics[width=3.2in,bb=21 -3 658 420]{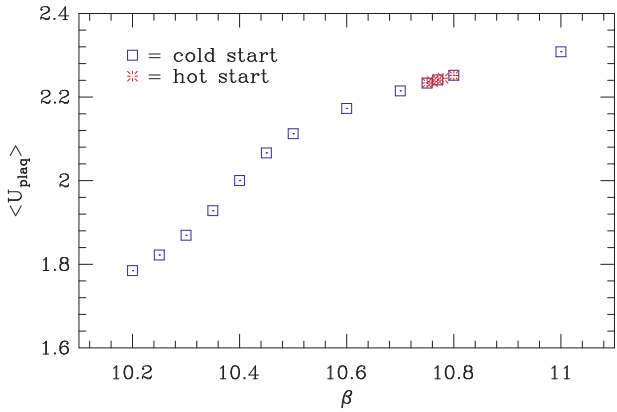}
\caption{Plaquette vs.\ $\beta$ on a $6\times 12^3$ lattice.
The normalization is such that $\langle U_\mathrm{plaq}\rangle = N_c$ 
in the free theory.}
\label{fig:plaq_126}
\end{figure}
In our $6\times 12^3$ calculations we find that the jump in
$\langle|L|\rangle$ is between $\beta=10.75$ and $\beta=10.80$
(Fig.~\ref{fig:plmag_126}).  This move in \(\beta\) of the critical
point is consistent with the conjecture that the deconfinement is a
thermodynamic phenomenon.  On the other hand the increase in the plaquette
is over the same $\beta$ (10.2-10.6) region as for
$N_t=4$ (see Fig.~\ref{fig:plaq_126}), as is expected for a bulk
transition.  And with \(N_t=6\) the bulk and finite-temperature phase
transitions are clearly separated.

\begin{figure}
\includegraphics[width=3.2in,bb=54 28 546 696]{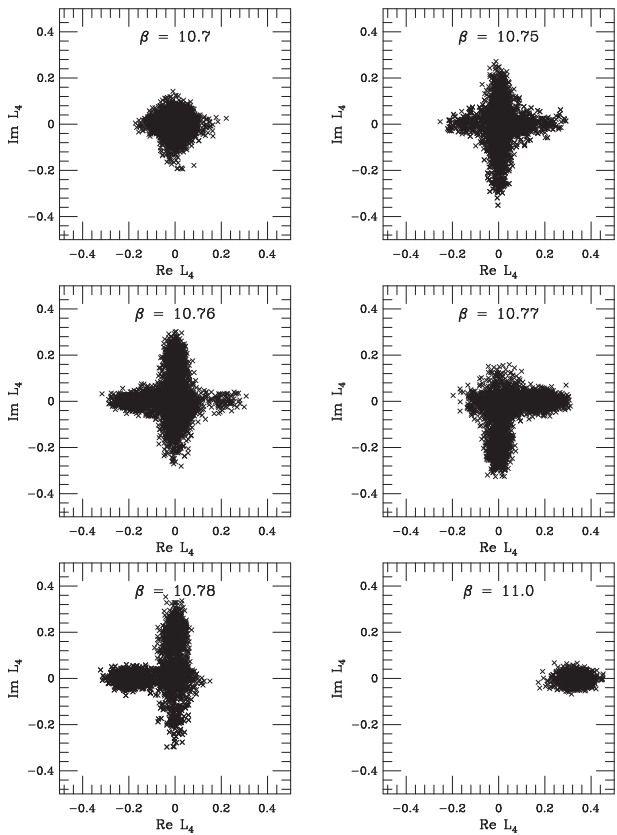}
\caption{Plots at six values of $\beta$
showing the phase and magnitude of the average
fundamental Polyakov loop on consecutive $6\times 12^3$ configurations.
Each cross corresponds to the value of $L_4$ computed on a single
configuration.
}
\label{fig:pl_126}
\end{figure}

The six plots
shown in Fig.~\ref{fig:pl_126} show the real and imaginary parts of the
Polyakov loop for the last 2000 sweeps of calculations at
the corresponding couplings.  One can see the spontaneous
breaking of the Z(4) symmetry as the deconfinement transition
is crossed.  

One quantitative estimation of the critical coupling, $\beta_c$,
comes from the deconfinement fraction~\cite{ref:CT}.
Let us define $\phi\in[0,\pi/4]$ as the angle between arg($L_4$) and the
nearest Z(4) symmetry axis.  Given another angle $\theta\in[0,\pi/4]$,
one counts the number of configurations where $\phi\le\theta$, $N_{in}$,
versus the number of configurations where $\phi>\theta$, $N_{out}$.
In the confined, Z(4)-symmetric phase, on average,
\begin{equation}
{N_{in}\over N_{in} + N_{out} } ~=~ { \theta \over (\pi/4) }.
\end{equation}
The deconfinement fraction is the excess number of configurations
which have $\phi\le\theta$:
\begin{equation}
f(\theta) ~\equiv~ {\pi/4 \over (\pi/4) - \theta}\bigg[
{N_{in}\over N_{in} + N_{out} } ~-~ { \theta \over (\pi/4) }\bigg],
\end{equation}
where the factor outside the brackets normalizes the totally deconfined
$f(\theta)$ to one.  (Note that due to statistical fluctuations, 
$f(\theta)$ can be slightly negative in the confined phase.)
The critical coupling is defined to be the value of $\beta$ for
which $f(\theta) = 1/2$.
\begin{figure}
\includegraphics[width=3.2in,bb=19 -3 658 413]{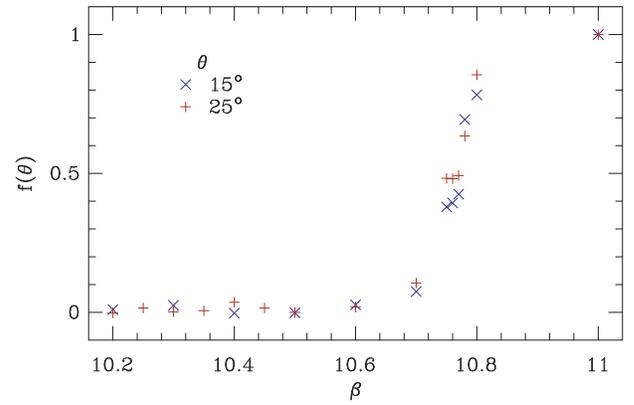}
\caption{Deconfinement fraction (defined in text) vs. $\beta$ 
for the $6\times 12^3$ lattice.
}
\label{fig:frac}
\end{figure}
In Fig.~\ref{fig:frac} we plot the deconfinement fraction 
with $\theta=15^\circ$ and $25^\circ$
for the $6\times 12^3$ lattice.  
We find $\beta_c = 10.78\pm 0.01$, where
the uncertainty is estimated by varying $\theta$ between 
$15^\circ$ and $30^\circ$.  

\begin{figure}
\includegraphics[width=3.2in,bb=19 -3 658 413]{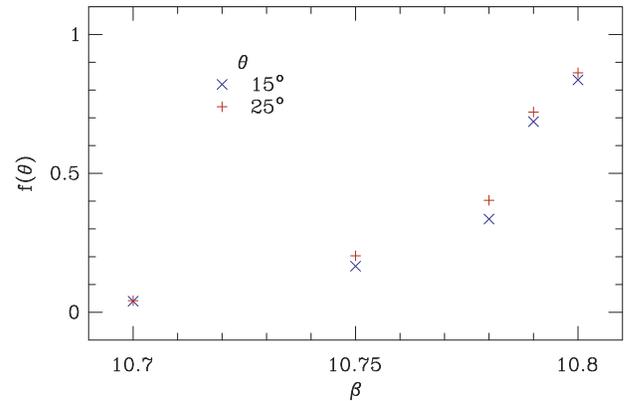}
\caption{Deconfinement fraction vs. $\beta$
for the $6\times 16^3$ lattice.}
\label{fig:frac_166}
\end{figure}
\begin{figure}
\includegraphics[width=3.2in,bb=8 -3 658 420]{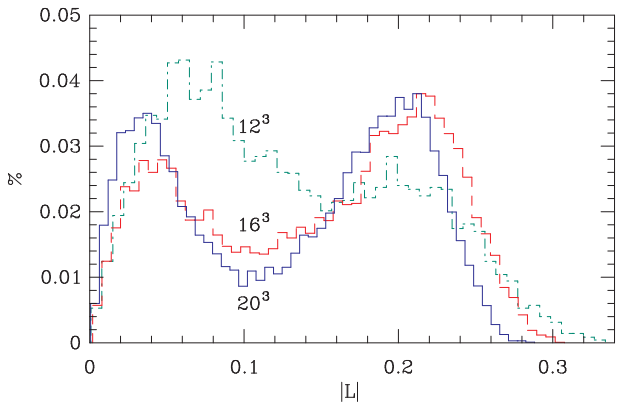}
\caption{Histogram of $|L_4|$ at $\beta_c = 10.78, 10.79, 10.79$ on
volumes of $12^3, 16^3, 20^3$, respectively.
}
\label{fig:hist_plmag_3vol}
\end{figure}

In order to determine the order of the phase transition,
we increased the spatial volume to $16^3$ and $20^3$.
With the larger volumes, the critical coupling increases
slightly to $\beta_c = 10.79$ as is expected (see Fig.\ref{fig:frac_166}).
The histograms of Polyakov loop magnitude \(|\langle L_4
\rangle |\) obtained from the larger two lattice volumes near their
respective critical points show two peaks, in clear contrast to the
\(12^3\) volume.  See Figure~\ref{fig:hist_plmag_3vol}.  This suggests a
first-order phase transition.

\begin{figure}
\includegraphics[width=3.2in,bb=19 1 595 699]{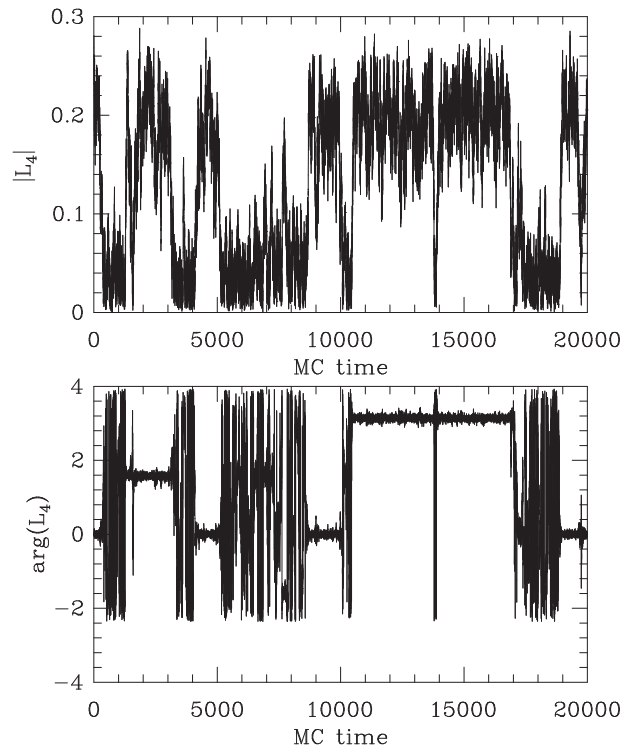}
\caption{Monte Carlo evolution of the magnitude (top) and the argument
(bottom) of the fundamental Polyakov loop for $\beta=10.79$ on the
$6\times 20^3$ volume.  We take \({\rm arg}(L_4)\in
[-\frac{3\pi}{4},\frac{5\pi}{4}]\) for clarity.}
\label{fig:histories}
\end{figure}

Indeed Polyakov loop evolution in simulation time, in
Figure~\ref{fig:histories}, signals coexistence of the confined and
deconfined phases at this temperature, \(\beta=10.79\).  The magnitude
stays with its low (confined) or high (deconfined) value for a relatively
long period, but occasionally jumps very quickly from one to the other
value.  And when the magnitude is low, the argument takes random
arbitrary values, while it is fixed to the neighborhood of one of the
four allowed Z(4) values when the magnitude is high.
\begin{figure}
\vspace{2.5in}
\includegraphics{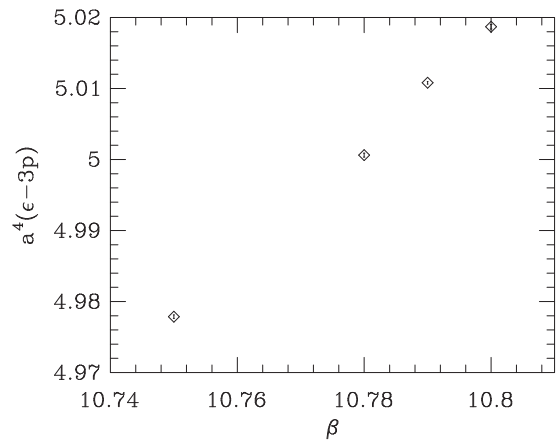}
\caption{Energy density minus 3 times the pressure vs.\ 
$\beta$ for the $6\times 16^3$ lattice. A divergent vacuum
contribution remains to be subtracted.}
\label{fig:em3p_166_001121}
\end{figure}
\begin{figure}
\vspace{2.5in}
\includegraphics{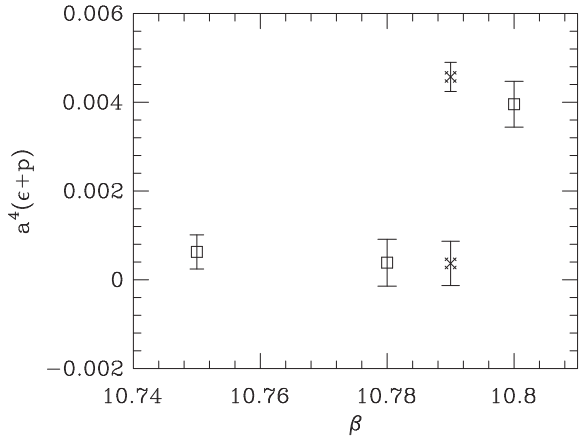}
\caption{Energy density plus pressure vs.\ $\beta$ for the
$6\times 16^3$ lattice.  Squares correspond
to averaging over the whole data set, the fancy crosses at the critical
beta correspond to separating into hot and cold phases as described
in the text.}
\label{fig:epp_166_001121}
\end{figure}

By combining these histogram and evolution observations, we conclude that
the finite-temperature deconfining phase transition of SU(4) Yang-Mills
system is of first-order.
%%% new latent heat (sub)section
It is thus desirable to compute the latent heat through combinations
of the energy density $\epsilon$ and the pressure $p$ \cite{Engels:1982qx}.  
Specifically we compute
\begin{eqnarray}
a^4(\epsilon - 3p) &=& - {6 N_c} a \frac{\partial g^{-2}}{\partial a}
( \ovl{P_t} + \ovl{P_s} )\\
a^4(\epsilon + p) &=& \frac{8 N_C}{g^2} C(g^2) (\ovl{P_t} - \ovl{P_s}) \, .
\end{eqnarray}
The average space-space and space-time (or space-temperature)
plaquettes 
are normalized such that if $g^2 = 0$, $\ovl{P_t} = \ovl{P_s} = 1$:
\begin{eqnarray}
\ovl{P_t} & = & \frac{1}{3\Omega}\frac{1}{N_c}\sum_x\sum_i 
{\rm Re~ Tr}_f P_{0i}(x) \\
\ovl{P_s} & = & \frac{1}{3\Omega}\frac{1}{N_c}\sum_x\sum_{i,j:i<j} 
{\rm Re~ Tr}_f P_{ij}(x)
\end{eqnarray}
where $\Omega$ is the $4d$ volume and $i$ and $j$ are spatial
indices.
In bare lattice perturbation theory the $\beta-$function
and Karsch coefficient are given, respectively, by \cite{Karsch:1982ve}
\begin{equation}
a\frac{\partial g^{-2}}{\partial a} = -2 \frac{11 N_c}{3(16\pi^2)}
+ O(g^2)
\label{eq:betafn}
\end{equation}
and 
\begin{eqnarray}
C(g^2) = 1 & - & \left[ 4N_c \left( \frac{N_c^2 -1}{32 N_c^2} 0.586844
- 0.005306\right)\right. \nonumber \\
&+& \left.\frac{11 N_c}{6(16\pi^2)}\right] g^2
\label{eq:karsch}
\end{eqnarray}
It is possible, and advisable, to use mean-field improved perturbation
theory or a nonperturbative calculation of these quantities for
an accurate calculation of the energy and pressure \cite{Boyd:1996bx}; 
however, for the purpose of establishing a nonzero latent heat, 
bare perturbation theory suffices.  

The quantities $\epsilon - 3p$ and $\epsilon + p$ are plotted
as functions of $\beta$ on the $16^3$ lattice and shown in
Figures \ref{fig:em3p_166_001121} and 
\ref{fig:epp_166_001121}, respectively.  The fancy crosses
in the latter figure correspond to separating the configurations
at $\beta_c = 10.79$ into hot and cold phases.  Note that 
$\epsilon - 3p$ plotted in Fig.\ \ref{fig:em3p_166_001121}
contains a divergent vacuum contribution which may be subtracted
after a zero temperature simulation is performed; however,
such subtraction is not necessary in order to compute the
latent heat from a discontinuity in $\epsilon - 3p$ at the
critical coupling $\beta_c$.
The separation of phases at $\beta_c$
was made on the basis of whether $|L_4|$ was greater or lesser
than some value $r$.  Based on the histograms in 
Fig.\ \ref{fig:hist_plmag_3vol} we varied $r$ from 0.08 to 0.14.
Table~\ref{tab:lh} lists the values for $\Delta(\epsilon - 3p)$ and
$\Delta(\epsilon + p)$ obtained for different $r$ on both
the $16^3$ and $20^3$ volumes.  The variation as a function of
$r$ is within the statistical errors.  

Thus, we observe a latent heat which is many standard deviations
greater than zero.  We also see that
$\Delta(\epsilon - 3p) \ne \Delta(\epsilon + p)$ which implies
a discontinuous change in pressure across the transition.  
If, for example, we take the $6\times20^3$ data with $r=0.10$,
we find (with statistical errors only)
\begin{eqnarray}
\Delta\epsilon &=& 5.7(3) T_c^4  \\
\Delta p &=& -0.45(13) T_c^4 \, .
\end{eqnarray}  
A nonzero $\Delta p$ was also seen in early studies of SU(3)
\cite{Svetitsky:1983bq,Brown:1988qe} and 
disappeared when going from the perturbative estimates for
(\ref{eq:betafn}) and (\ref{eq:karsch}) to nonperturbative
calculations \cite{Boyd:1996bx}.

\begin{table}
\caption{Discontinuities in $\epsilon-3p$ and $\epsilon+p$
at $\beta = 10.79$ for various values of $r$ (described in 
text).  The lattice spacing has been set through the critical
temperature: $a(\beta_c) = (N_t T_c)^{-1}$.}
\begin{center}
\begin{tabular}{cccc}
\hline\hline
$\Omega$ & $r$ & $\Delta(\epsilon-3p)/T_c^4$ & 
$\Delta(\epsilon+p)/T_c^4$ \\
\hline
%$6\times16^3$  & 0.08 & ~$5.8(3)\times10^{-3}$~  & ~$3.7(6)\times10^{-3}$~ \\
%        & 0.10 & ~$5.9(3)\times10^{-3}$~  & ~$4.2(6)\times10^{-3}$~ \\
%        & 0.12 & ~$5.9(3)\times10^{-3}$~  & ~$4.1(6)\times10^{-3}$~ \\ 
%        & 0.14 & ~$5.9(3)\times10^{-3}$~  & ~$4.4(6)\times10^{-3}$~ \\ \hline
%$6\times20^3$  & 0.08 & ~$5.2(2)\times10^{-3}$~  & ~$3.8(3)\times10^{-3}$~ \\
%        & 0.10 & ~$5.4(2)\times10^{-3}$~  & ~$4.0(3)\times10^{-3}$~ \\
%        & 0.12 & ~$5.4(2)\times10^{-3}$~  & ~$4.1(3)\times10^{-3}$~ \\
%        & 0.14 & ~$5.5(2)\times10^{-3}$~  & ~$4.2(3)\times10^{-3}$~ \\
$6\times16^3$  & 0.08 & ~$7.5(4)$~  & ~$4.8(8)$~ \\
        & 0.10 & ~$7.6(4)$~  & ~$5.4(8)$~ \\
        & 0.12 & ~$7.6(4)$~  & ~$5.3(8)$~ \\ 
        & 0.14 & ~$7.6(4)$~  & ~$5.7(8)$~ \\ \hline
$6\times20^3$  & 0.08 & ~$6.7(3)$~  & ~$4.9(4)$~ \\
        & 0.10 & ~$7.0(3)$~  & ~$5.2(4)$~ \\
        & 0.12 & ~$7.0(3)$~  & ~$5.3(4)$~ \\
        & 0.14 & ~$7.1(3)$~  & ~$5.4(4)$~ \\
\hline\hline
\end{tabular}
\label{tab:lh}
\end{center}
\end{table}

A thorough calculation of the latent heat in the
SU(4) deconfinement transition requires a full study
of the lattice spacing dependence as well as nonperturbative
determination of the $\beta$-function and Karsch coefficient.
However, even the exploratory study here makes clear
the latent heat is nonzero and further establishes 
the first-order nature of the phase transition. 

We can compare the latent heat for SU(4) to that for SU(3)
by normalizing by the energy density for an ideal
gluon gas.  If we take the latent heat to be
$\Delta \epsilon/T_c = 6.0 \pm 1.5$ and divide by the
Stefan--Boltzmann energy density,
\begin{equation}
\epsilon_{\rm SB}(T) ~=~ \frac{(N_c^2 - 1)\pi^2}{15} T^4
\end{equation}
we find
\begin{equation}
\frac{\Delta \epsilon}{\epsilon_{\rm SB}(T_c)} ~=~ 0.60 \pm 0.15 \, .
\end{equation}
Our result should be compared against
the $N_t = 6$ SU(3) latent heat obtained using a perturbative
$\beta-$function: $\Delta \epsilon/\epsilon_{\rm SB}=0.454(11)$
\cite{Beinlich:1997xg}.
A state-of-the-art SU(3) calculation, which used
an improved action and a nonperturbative $\beta-$function,
gave $\Delta \epsilon/\epsilon_{\rm SB}= 0.266(17)$ 
\cite{Beinlich:1997xg}.  Further work is required to see if the
effect of going from a perturbative to nonperturbative 
$\beta-$function is as dramatic for SU(4) as for SU(3).

%%%%%%
\section{String tensions}
\label{sec:string}

\begin{figure}
\includegraphics[width=3.2in,bb=86 66 537 526]{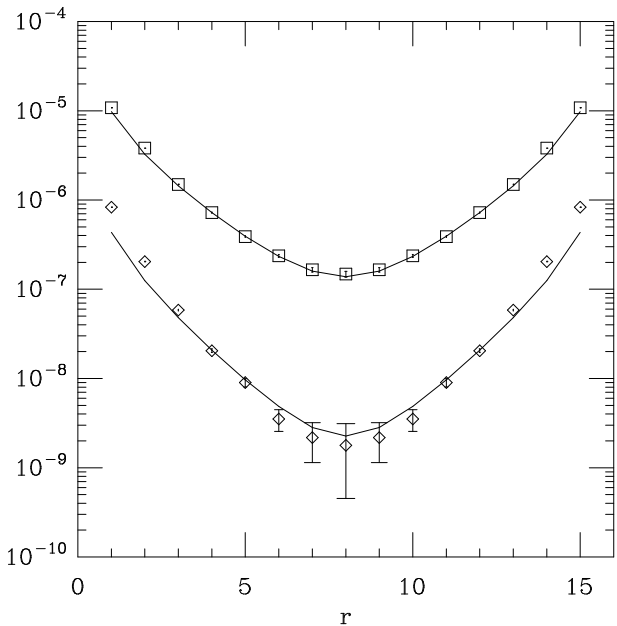}
\caption{Polyakov loop correlation function in
{\bf 4} (top) and {\bf 6} (bottom) representations on a
$6\times 16^3$ lattice at $\beta=10.65$.  The symbols are the data
points and the solid lines are fits in the range \(4 \le r \le 12\).
}
\label{fig:su4_0616_b1065_PcorrR}
\end{figure}
\begin{figure}
\includegraphics[width=3.2in,bb=92 66 537 526]{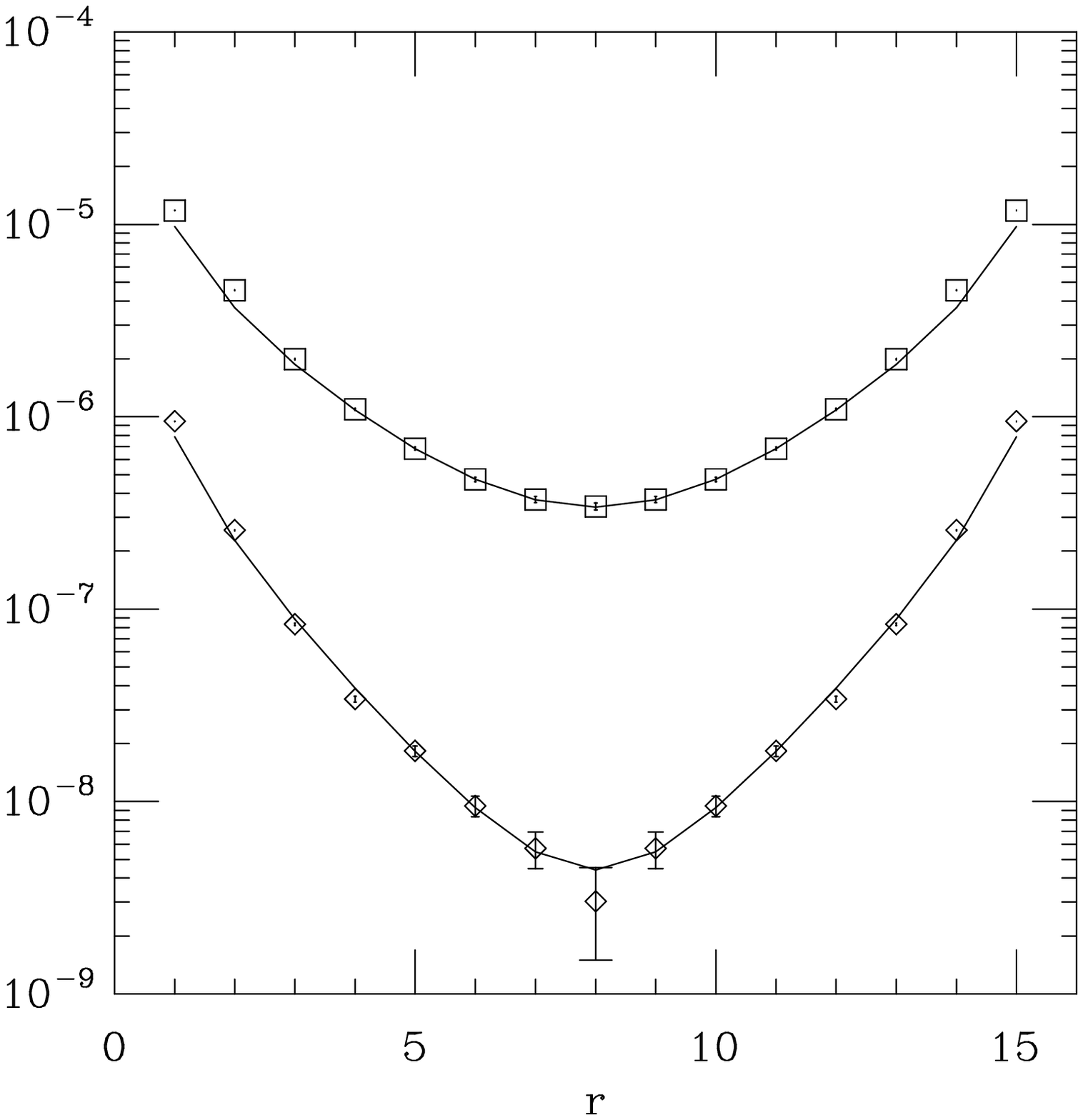}
\caption{Polyakov loop correlation function in
{\bf 4} (top) and {\bf 6} (bottom) representations on a
$6\times 16^3$ lattice at $\beta=10.70$.  The symbols are the data
points and the solid lines are fits in the range \(5 \le r \le 11\).
}
\label{fig:su4_0616_b1070_PcorrR}
\end{figure}
\begin{table}[b]
\caption{String tensions (in lattice units) between static
fundamental (\(k=1\)) and diquark (\(k=2\)) charges, and their ratio,
using every \((N_{\rm skip}+1)\)th configuration in the analysis.
The quoted uncertainties are statistical.}
\begin{center}
\begin{tabular}{rcrrr}
\hline\hline
\multicolumn{1}{c}{\(\beta\)}&
\multicolumn{1}{c}{\(N_{\rm skip}\)}&
\multicolumn{1}{c}{\(\sigma_1\)}&
\multicolumn{1}{c}{\(\sigma_2\)}&
\multicolumn{1}{c}{\(\sigma_2/\sigma_1\)}
\\
\hline
            & 0\hfill & 0.098(2)\hfill & 0.138(14)\hfill & 1.45(15)\hfill \\
10.65\hfill & 4\hfill & 0.092(4)\hfill & 0.137(30)\hfill & 1.67(36)\hfill \\
	    & 9\hfill & 0.101(7)\hfill & 0.164(56)\hfill & 1.77(67)\hfill \\
\hline
            & 0\hfill & 0.076(2)\hfill & 0.118(13)\hfill & 1.59(13)\hfill \\
10.70\hfill & 4\hfill & 0.080(4)\hfill & 0.154(40)\hfill & 1.91(49)\hfill \\
            & 9\hfill & 0.084(6)\hfill & 0.163(53)\hfill & 2.03(68)\hfill\\
\hline\hline
\end{tabular}
\label{tab:tensions}
\end{center}
\end{table}
We use two different lattices, $6\times 16^3$ and $8\times 12^3$,
for studying string tensions.
For the former, we choose the coupling values of
\(\beta\)=10.65 and 10.70, safely away from both the bulk and the
deconfining phase transitions, and in the confining phase (see
Figure~\ref{fig:pl_126}).  Polyakov loop correlations 
(see Eq.\ (\ref{eq:corrfn})) for the fundamental
({\bf 4}, \(k=1\), top) and anti-symmetric diquark ({\bf 6}, \(k=2\),
bottom) representations are shown in
Figures~\ref{fig:su4_0616_b1065_PcorrR} and \ref{fig:su4_0616_b1070_PcorrR}. 
A clear difference in the rates of exponential decay is observed
between \(C_{\bf 4}\) and \(C_{\bf 6}\). 
Using a correlated, jackknifed fit to the form~\cite{ref:FSST}
\begin{equation}
\frac{a_k}{r}\exp[-V_k(r)N_t] + \frac{a_k}{N_s-r}\exp[-V_k(N_s-r)N_t],
\end{equation}
with \(N_t=6\), \(N_s=16\) and
\begin{equation}
V_k(r) = \sigma_k r - \frac{\pi r}{3 N_t^2}
\end{equation} 
we obtain string tensions, \(\sigma_1\) and \(\sigma_2\), and their ratio, 
tabulated in Table~\ref{tab:tensions}.
The analysis of the correlation
functions is done using every measurement, every fifth measurement,
and every tenth measurement in order to estimate correlations between
successive measurements (each separated by 10 Monte Carlo steps,
see Sec.\ \ref{sec:simul}).  The increase in the statistical error 
with the number of skipped configurations, $N_{\rm skip}$, indicates
a significant auto-correlation.  Unfortunately, it appears that several
hundred configurations are necessary in order to obtain a precise fit,
so we cannot drop too many of the measurements.
However, we can infer from our data that
\begin{equation}
\frac{\sigma_2}{\sigma_1} > 1 \, ,
\end{equation}
by roughly 2 standard deviations.
Note that both \(\sigma_1\) and \(\sigma_2\) decrease as 
\(\beta \to \beta_c\) (i.e.\ as \(T\) increases).  Since the
lattice spacing decreases as \(T\) increases, the fit range
for \(\beta=10.70\) does not include the $r=4$ and $r=12$ data
(see captions of Figs.\ \ref{fig:su4_0616_b1065_PcorrR} and 
\ref{fig:su4_0616_b1070_PcorrR}).
Our numerical accuracy is good enough to conclude there are
two different strings, one between the fundamental charges carrying one
unit of flux, and another, stronger, between the diquark charges
carrying two units of flux.  It is not yet good enough, however, to
distinguish among various predictions for this ratio summarized by
Strassler \cite{ref:STRASSLER}.  However, this establishes
numerically the expectation for \(\sigma_1 \ne \sigma_2\)
in SU(4) Yang-Mills theory, just as Ref.\ \cite{Ohta:1986pc} showed 
\(\sigma_1 = \sigma_2\) in SU(3) Yang-Mills theory.

A string model~\cite{ref:PISARSKI_ALVAREZ} predicts that 
\begin{equation}
\frac{T_c}{\sqrt{\sigma_1(T=0)}}
\approx \sqrt{\frac{3}{\pi(d-2)}} = 0.69 \, ,
\label{eq:string}
\end{equation}
which is quite close for SU(3)~\cite{ref:KARSCH_LAT99}.
We have not computed the zero temperature string tension,
but only the string tension roughly near \(T_c\), to find
\begin{equation}
\frac{T_c}{\sqrt{\sigma_1(T\approx T_c)}}
= 0.60 \, .
\end{equation}
The extent which the lattice scale changes between
\(\beta = 10.70\) and \(\beta_c = 10.79\) is main uncertainty
above.  Of course a zero temperature study is necessary before
one can assess the agreement with Eq.\ (\ref{eq:string}).
 
\begin{figure}
\includegraphics[width=3.2in,bb=69 55 334 296]{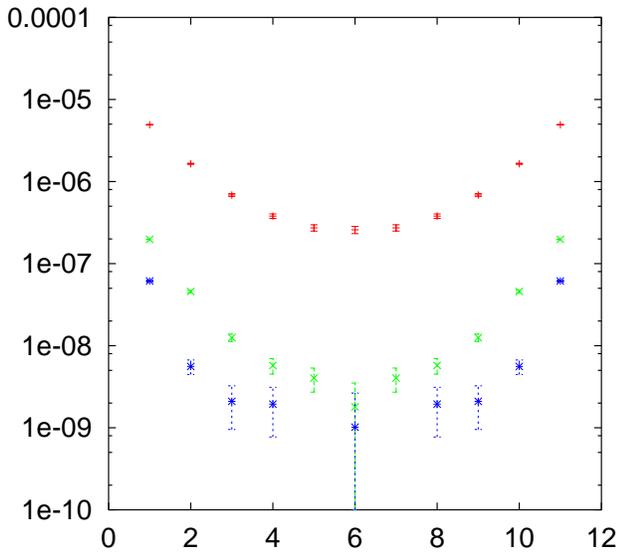}
\caption{Polyakov loop correlation function in
{\bf 4} (top), {\bf 6} (middle), and {\bf 15} (bottom) 
representations on a $12^3\times 8$ lattice at $\beta=10.85$
}
\label{fig:su4_0812_b1085_Pcorr}
\end{figure}

On this lattice of \(6\times 16^3\), which is coarser and larger
of the two, no signal was obtained for either the symmetric diquark ({\bf
10}) or adjoint ({\bf 15}) representations.  In contrast, with the finer
lattice spacing (at \(\beta=10.85\)) on the smaller \(8\times 12^3\)
lattice, flattening of the adjoint correlation is observed (see
Figure~\ref{fig:su4_0812_b1085_Pcorr}).   This suggests the breaking of
confining string for the adjoint representation at a rather short
distance of 3 lattice spacings.  It gives us
confidence that the correlations on the \(6\times 16^3\) lattice
should be dominated by the non-perturbative strings for ranges longer
than at least 3 lattice units.  Notice also that while
string breaking is an expected behavior for the adjoint representations in
general \cite{ref:STRINGBREAKING}, such an absence of
string is yet to be observed in SU(3) Yang-Mills theory which employs much
finer and larger lattices than the present work.  

%%%%%%
\section{Conclusions}
\label{sec:concl}

We have revisited the confinement--deconfinement transition of SU(4)
Yang-Mills theory through Monte Carlo lattice calculation.  One problem
with the earlier results is that the deconfinement transition with $N_t =
4$ is very close in coupling constant space to a known bulk transition,
so that its finite-temperature nature or its order is not clear.  We have
shown that by decreasing the lattice spacing by $2/3$, the deconfinement
transition moves upward in the coupling and proves itself as a
finite-temperature transition, and it becomes well-separated from the
bulk transition which does not move.  Nevertheless, we observe a clear
signal for coexistence of confined and deconfined phases at this
deconfinement transition.  Therefore, we confirm that the deconfinement
transition of SU(4) Yang-Mills theory is first-order.
Additionally a first calculation of the latent heat of the
SU(4) deconfinement transition has been presented here, giving
$\Delta\epsilon \approx 6 T_c$, or $\Delta\epsilon /\epsilon_{\rm SB}
\approx 0.6$.  Using improved techniques, the SU(3) latent heat
is $\Delta\epsilon /\epsilon_{\rm SB} = 0.266(17)$ \cite{Beinlich:1997xg}, 
and it will be interesting to see how the latent heat depends on
$N_c$.

Our calculations of the string tensions are a first study
in lattice SU(4) and should be improved to meet the current
state-of-the-art which exists for SU(3).  Even so, we
observe a ratio for {\bf 4} and {\bf 6} dimensional string
tensions which is between 1 and 2.  It also appears that
the adjoint string breaks at a short distance.  We hope this work
shows that it is interesting and feasible to study ratios of
string tensions for $N_c > 3$ lattice simulations.

%%%%%%
\section*{Acknowledgments}

We are indebted to the MILC collaboration~\cite{ref:MILC_CODE}
whose pure-gauge SU(3) code was adapted for this work. 
The majority of our calculations were performed on
a cluster of Pentium III processors in the BNL Computing Facility.
We acknowledge helpful conversations with M.\ Creutz,
R.\ Pisarski, and M.\ Strassler. 
Thanks also to RIKEN, Brookhaven National Laboratory, 
and the U.S. Department of Energy for
providing the facilities essential for the completion of this work.

%%%%%%%%%%%

%%% figure 
%\begin{figure}
%\vspace{7.5in}
%\special{psfile=plmag_hist_206_b1079.eps hscale=75 vscale=75 
%hoffset=0 voffset=0}
%\caption{Monte Carlo evolution (top) and histogram (bottom)
%of the fundamental Polyakov loop magnitude for $\beta=10.79$ on
%the $20^3 \times 6$ volume.
%}
%\label{fig:plmag_hist_206_b1079}
%% FIGURE_FILE plmag_hist_206_b1079.eps
%\includegraphics[width=3.2in,bb=0 0 576 576]{plmag_hist_206_b1079.eps}
%\end{figure}

%% figure 
%\begin{figure}
%\vspace{7.5in}
%\special{psfile=plarg_hist_206_b1079.eps hscale=75 vscale=75 
%hoffset=0 voffset=0}
%\caption{Monte Carlo evolution (top) and histogram (bottom)
%of the argument of the fundamental Polyakov loop for $\beta=10.79$ on
%the $20^3 \times 6$ volume.
%}
%\label{fig:plarg_hist_206_b1079}
%% FIGURE_FILE plarg_hist_206_b1079.eps
%\includegraphics[width=3.2in,bb=0 0 576 576]{plarg_hist_206_b1079.eps}
%\end{figure}
\end{document}